\documentclass[aps,prl,superscriptaddress,twocolumn,showpacs]{revtex4}

\renewcommand*{\[}{\begin{equation}}

\renewcommand*{\]}{\end{equation}}

\def\PRA{{Phys.~Rev.~A} }

\def\JPB{{J.~Phys.~B} }
\def\PRL{{Phys.~Rev.~Lett.} }

\newcommand{\myscaleboxa}[1]{\scalebox{0.3}[0.3]{#1}}
\newcommand{\myscaleboxb}[1]{\scalebox{0.45}[0.45]{#1}}

\usepackage{epsfig}

\usepackage{hyperref}

\begin{document}

\title{Origin of Plateau and Species dependence of Laser-Induced High-Energy Photoelectron Spectra}

\author{Zhangjin Chen}

\affiliation{J. R. Macdonald Laboratory, Physics Department, Kansas
State University, Manhattan, Kansas 66506-2604, USA}

\author{Anh-Thu Le}

\affiliation{J. R. Macdonald Laboratory, Physics Department, Kansas
State University, Manhattan, Kansas 66506-2604, USA}

\author{Toru Morishita}

\affiliation{Department of Applied Physics and Chemistry, University
of Electro-Communications, 1-5-1 Chofu-ga-oka, Chofu-shi, Tokyo,
182-8585, Japan and PRESTO, JST Agency, Kawaguchi, Saitama 332-0012,
Japan}

\author{C. D. Lin}

\affiliation{J. R. Macdonald Laboratory, Physics Department, Kansas
State University, Manhattan, Kansas 66506-2604, USA}

\date{\today}

\begin{abstract}

We analyzed the energy and momentum distributions of laser-induced
high-energy photoelectrons of alkali and rare gas atoms. For the
plateau electrons with energies above $4U_p$, ($U_p$ is the
ponderomotive energy), in the tunneling ionization regime, we showed
that they originate from the backscattering of laser-induced
returning electrons. Using the differential elastic scattering cross
sections between the target ion with \emph{free} electrons, we
explain experimental observations of whether the plateau electron
spectra is flat or steeply descending, and their dependence on
species and laser intensity. This quantitative rescattering theory
can be used to obtain energy and momentum distributions of plateau
electrons without the need of solving the time-dependent
Schr\"{o}dinger equation, but with similar accuracy.

\end{abstract}


\maketitle

When atoms or molecules are placed in an intense laser pulse, an
electron can be released through either a multiphoton or a tunneling
mechanism. The distinction is based on the Keldysh parameter
$\gamma=\sqrt{I_p/2U_p}$, where $I_p$ is the ionization energy and
$U_p=A_0^2/4$ is the ponderomotive energy with $A_0$ is the peak
value of the vector potential. In the multiphoton regime
($\gamma>1$), the electron spectra exhibit characteristic
above-threshold ionization (ATI) peaks separated by photon energy,
with yields decreasing monotonically with increasing electron energy
\cite{Agostini}. At higher intensities, in the tunneling region
($\gamma<1$), the spectra are notably different. First, the
ionization yield drops steeply from the threshold, but from about 3
or $4U_p$ onward, the yield flattens out significantly until at
about $10U_p$ where it drops precipitously again. The flattened
spectral region from 4-$10 U_p$ is called the plateau. Similar
plateau and cutoff are also well-known in the high-order harmonic
generation (HHG). Despite this canonical description, the HHG and
electron spectra in the plateau region are not always flat. For
electron spectra, earlier experiments show pronounced enhancement in
the plateau region in potassium, but not in sodium \cite{Gaarde}.
Similarly, using nearly identical lasers, clear plateau shows up in
Xe target, but not in Kr and Ar \cite{Paulus_PRL94}. Experimentally
the energy spectra of plateau electrons have been observed to depend
on laser intensities \cite{Grasbon}. The origin of plateau electrons
and their dependence on target species and laser intensity is
explained in this Letter.

Considerable understanding of laser-induced ATI spectra from atoms
has been achieved since 1990's \cite{yang93,Paulus_JPB94,Walker}.
While the observed spectra in general can be reproduced from solving
the time-dependent Schr\"{o}dinger equation (TDSE) within the single
active electron approximation, interpretation of the plateau
electrons is based on the rescattering model. In this model
electrons that are released earlier by tunneling ionization can be
driven back by the laser field to recollide with the target ion. The
plateau electrons are understood to originate from the elastic back
scattering of the returning electrons by the target ion. However,
existing rescattering model does not predict how they depend on
target species and laser intensity. In this Letter we provide such a
quantitative rescattering theory (QRS) where the energy and momentum
spectra of the plateau electrons can be directly calculated using
the elastic scattering cross sections between the target ion and
\emph{free} electrons, in combination with laser-induced returning
electron wave packet. We also show that the returning electron wave
packets depend largely on the laser parameters only. Thus the
behavior of the plateau electron spectra is determined
\emph{entirely} by the energy and species dependence of the elastic
scattering cross sections between the returning electrons and the
target ion.

To understand  electron energy spectra, we first examine how
electron-ion collisions contribute to the photoelectron momentum
distributions. In Fig.~1 we show a typical two-dimensional (2D)
electron momentum spectra for an atomic target calculated by solving
the TDSE \cite{chen06,toru07}. (We use linearly polarized electric
field for the laser pulse along the $z$ axis with the
carrier-envelope phase set to zero \cite{chen06}). The horizontal
axis is the direction of laser's polarization and the vertical axis
is any direction perpendicular to it (the electron spectra has
cylindrical symmetry). Due to the short laser pulse, the right-hand
side ($p_z>0$) and the left-hand side ($p_z<0$) spectra are not
symmetric. In Fig.~1, four semi-circles with photoelectron energies
of 2, 4, 6 and $10U_p$, respectively, are shown. On the right-hand
side, three other semi-circles are given, each with its center
shifted along the $p_z$ axis. Each of this circle can be expressed
vectorially by $\textbf{p}_{\nu}=-A_r\hat{\textbf{p}}_z+
p_r\hat{\textbf{p}}$. Measured from its own center, the circle maps
out the momentum space of the elastically scattered electron with
incident momentum $p_r$. The center is shifted since collision
occurs in the presence of the laser field. If the electron-ion
collision occurs at time $t_r$ when the vector potential is
$\textbf{A}(t_r)=A_r\hat{\textbf{p}}_z$, the electron will reach the
detector gaining an additional momentum $-A_r$. (Atomic units are
used in the above expressions such that the momentum gain in the
laser field is directly given by the vector potential $A_r$.) In the
figure, we have chosen the ``incident'' electrons to be from the
right, i.e., with negative $p_z$. The direction of
$\hat{\textbf{p}}$ and the elastic scattering angles $\theta$ are
measured from this ``incident'' direction. This choice would result
in high-energy or plateau electrons to emerge with positive $p_z$
after the electron undergoes large-angle backscatterings
($\theta>90^\circ$). Note that electrons scattered into the forward
directions will emerge with low energies. Clearly, similar circles
can be drawn for electrons ``incident'' from the left.

\begin{figure}
\mbox{\rotatebox{270}{\myscaleboxa{
\includegraphics{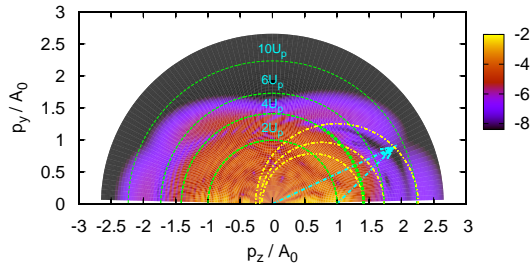}}}}
\caption{Typical 2D electron momentum distributions. Photoelectrons
with constant energy are given by concentric circles from the
origin. The three half circles on the right-hand side depict
momentum surfaces resulting from the elastic scattering of returning
electrons by the target ion in the presence of a laser field.
High-energy electrons are obtained only via large angles
backscattering.}
\end{figure}

According to the classical model, electrons that return to the ion
core at the time when the vector potential is at the peak, $A_0$, it
will have maximum kinetic energy of $3.17U_p=p_r^2/2$, i.e., with
$p_r=1.26A_0$.  If the electron is backscattered by $180^\circ$, the
emerging photoelectron would have a momentum of $2.26A_0$, or energy
of $10U_p$. In a recent paper \cite{toru08}, we examined the
photoelectron distributions after the $p_r=1.26A_0$ electrons have
been elastically scattered into the backward directions. The
electron yield along this ring (called BRR, or back rescattering
ridge) has been shown to be given by
\begin{eqnarray}
\label{BRR}I(p_\nu,\theta_\nu)=S(p_r)\sigma(p_r,\theta).
\end{eqnarray}
This equation is based on the concept of rescattering theory. It
states that photoelectron yields along the BRR are proportional to
the elastic differential cross section (DCS), $\sigma(p_r,\theta)$,
multiplied by a returning electron wave packet, represented by
$S(p_r)$, with the target ion. The validity of Eq.~(\ref{BRR}) was
established in \cite {toru08} based on accurate TDSE results. Its
validity has also been confirmed experimentally in two recent
reports \cite{Japan_08,USA_08}.

To examine the plateau electrons, elastic scattering by electrons
with returning energy less than $3.17U_p$ must be included. These
are electrons that return to the parent ion when $A_r=A(t_r)$ is not
at the peak. We set the returning electron's momentum $p_r=1.26A_r$
and using the rescattering model, Eq.~(\ref{BRR}). The validity of
this model can be checked by comparing the extracted
$\sigma(p_r,\theta)$ from the calculated photoelectron momentum
spectra $I(p_\nu,\theta_\nu)$, with those directly calculated from
electron-ion collisions, as carried out in Morishita \emph{et al}.
\cite{toru08}. Alternatively, we can first use Eq.~(\ref{BRR}) at an
arbitrary angle, say, for photoelectron with momentum vector
10$^\circ$ from the polarization axis, to derive the wave packet
$S(p_r)$. We then obtain the rescattering ``generated''
photoelectron spectra by using Eq.~(\ref{BRR}). This way we have a
quantitative rescattering theory which can be used to obtain the
momentum and energy spectra for plateau electrons. We remark later
that $S(p_r)$ can also be obtained by other approximations.

\begin{figure}
\mbox{\rotatebox{270}{\myscaleboxa{
\includegraphics{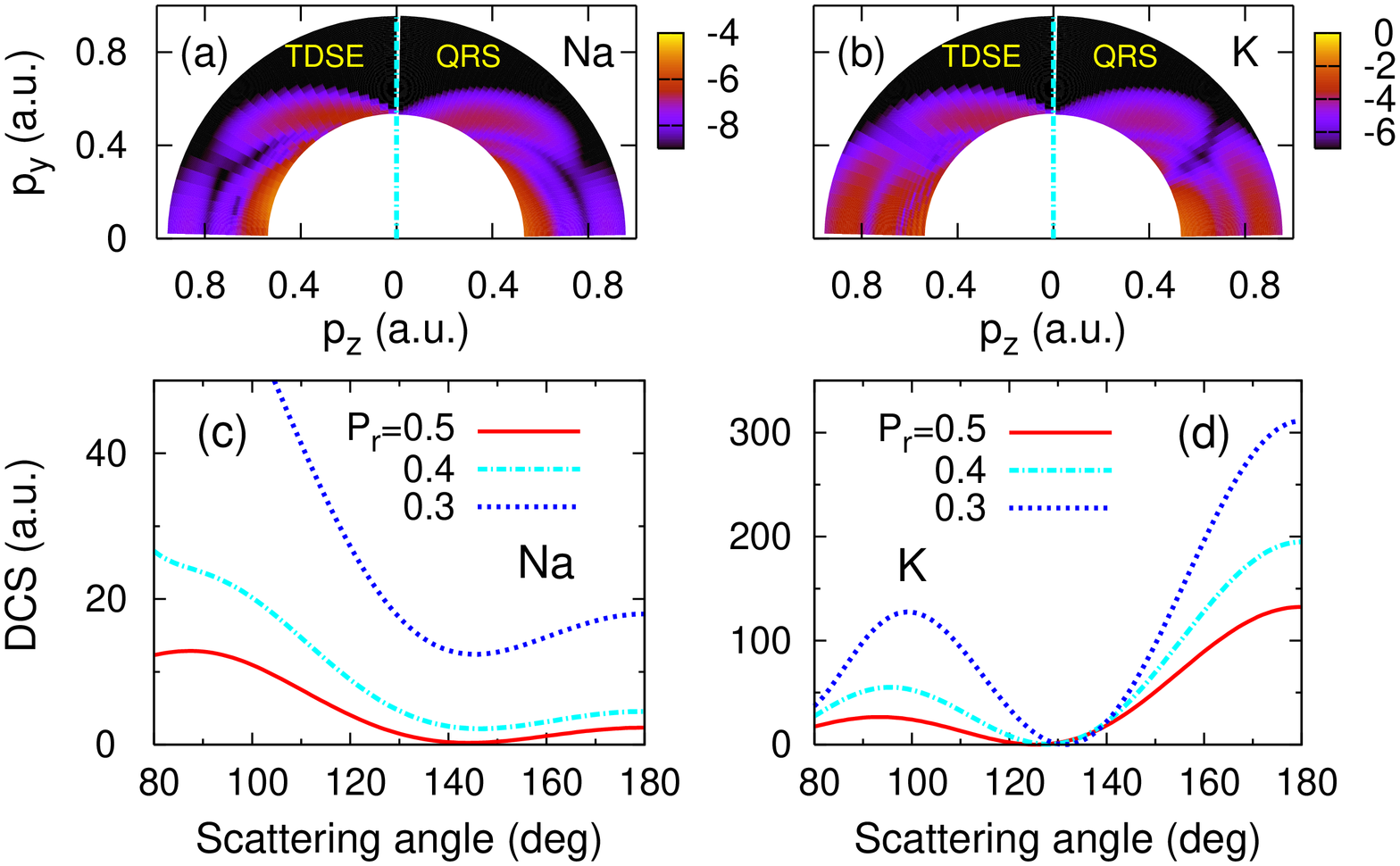}}}}
\mbox{\rotatebox{270}{\myscaleboxa{
\includegraphics{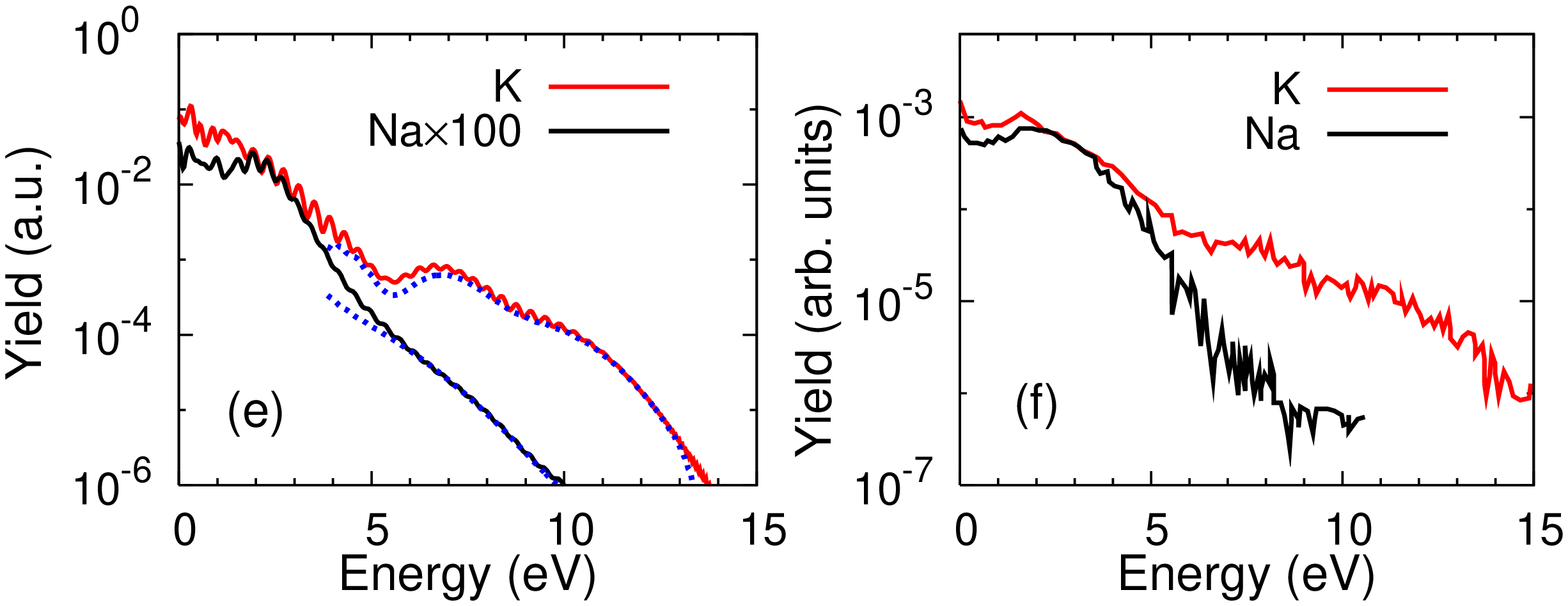}}}}
\caption{Single ionization of Na and K atoms by a 5-cycle, 3.2
$\mu$m infrared laser with peak intensity of $1.0\times 10^{12}$
W/cm$^{2}$ ($U_p=0.96$ eV). (a,b) Comparison of 2D momentum
distributions above $4U_p$ and $p_z\geq0$ (i.e., half of the
momentum space only) obtained from TDSE with those from QRS model.
Agreement between the two calculations is indicated by the good
reflection symmetry in each figure; (c,d) Elastic differential
electron-ion scattering cross sections at large angles; (e) Electron
energy spectra calculated from TDSE and those from the QRS model
(dotted lines) above $4U_p$ for each target; (f) Experimental
electron spectra taken from Gaarde \emph{et al}. \cite{Gaarde}. In
(e,f) the spectra from the two targets are normalized to each other
at low energies.}
\end{figure}

In Figs. 2(a,b) electron momentum spectra ($p_z\geq0$) for sodium
and potassium atoms exposed to a 5-cycle laser pulse with wavelength
of 3200 nm and peak intensity of 1~TW/cm$^2$ are presented. On the
left-hand half of the 2D plot, the results calculated from TDSE are
shown, and on the right-hand half the same distributions obtained
from the present QRS model are presented. If the two calculations
agree well, then there should exist a good reflection symmetry in
the figure. This is clearly the case for each target, confirming the
validity of the QRS model at the level of electron momentum
distributions for the plateau electrons.  In Figs.~2(c,d) the DCS
for e-Na$^+$ and e-K$^+$ collisions are shown for the range of
momenta $p_r$ that contribute to the 4-$10U_p$ photoelectrons. As
the collision energy increases, the DCS decreases. However, for the
same $p_r$, the DCS for e-K$^+$ is about 100 times larger.
Furthermore, the angular dependence of the DCS in Fig.~2(c,d) are
clearly reflected in the momentum distributions in Figs.~2(a,b). In
other words, the elastic scattering cross sections between free
electrons and target ion can be ``read'' out directly from the ATI
electron spectra in the momentum space. This is possible because of
the other important feature of QRS: the returning electron wave
packet $S(p_r)$ depends mostly on the lasers only. This fact has
been noted in the high energy region before \cite{chen07}.

From the momentum spectra the ATI energy spectra can be calculated.
Since the right-hand side and left-hand side 2D momentum spectra are
not symmetric for short pulses, we calculate the energy spectra from
each side separately and the sum of them reproduces the whole energy
spectra in which the left-hand side makes more contribution to the
lower energy part. In Fig.~2(e), the results from TDSE and from QRS
are compared. We limit the QRS model to electron energy above $4U_p$
only. Clearly the QRS model reproduces the TDSE results well. These
calculations can also be compared to the experimental data reported
in Gaarde \emph{et al}. \cite{Gaarde}, as shown in Fig. 2(f). There
is a close similarity in the relative yields between our
calculations and experiment, even though our calculations used a
5-cycle (20~fs FWHM) pulse while the experiment used somewhat
different intensities for a pulse with duration of 1.9~ps. The QRS
results clearly establish that the experimental plateau electron
spectra for Na and K are due to their elastic scattering cross
sections in the momentum range of 0.3 to 0.5. Although it has long
been understood (or speculated) that the behavior of plateau
electrons are related to elastic scattering cross sections
\cite{Walker,Gaarde}, no direct quantitative connection between the
two has ever been established until now.

As a side note, we comment that the lower end of the plateau
electrons has been set at $4U_p$ in this Letter. According to the
classical theory, direct ionization by tunneling (without
rescattering) will give maximum electron energy of $2U_p$. Between 2
and $4U_p$, the direct ionization part could still interfere with
the rescattering part \cite{chen07}, thus we chose $4U_p$ as the
lower end of the QRS model.

\begin{figure}
\mbox{\rotatebox{0}{\myscaleboxb{
\includegraphics{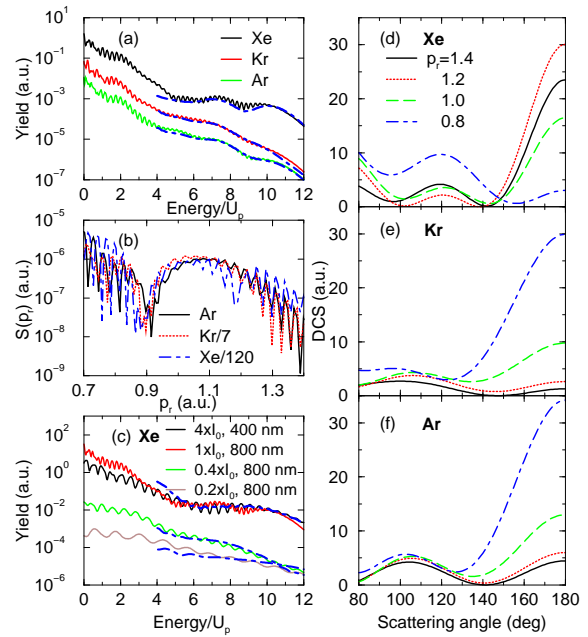}}}}
\caption{Single ionization of Xe, Kr and Ar atoms by a 5-cycle laser
pulse. Except for (c), laser's wavelength is 800 nm and the peak
intensity is $I_0=1.0\times10^{14}$ W/cm$^{2}$ ($A_0=0.94$). (a)
Comparison of electron energy spectra from TDSE, and from the QRS
model above $4U_p$ (chain lines). The QRS energy spectra for Kr and
Xe are calculated using the wave packet extracted from Ar, with
proper normalization at one energy point; (b) Normalized momentum
distributions of the returning electron wave packets extracted from
the ``left-side'' of the momentum spectra; (c) Comparison of Xe
plateau electron spectra at different laser intensities and
wavelengths, with laser parameters indicated in the figure. Note
that the QRS model fails at the lowest intensity; (d,e,f)
Differential electron-ion elastic scattering cross sections used in
the QRS model.}
\end{figure}

We have also studied the plateau electrons for Ar, Kr and Xe atoms.
Experimental data for these systems have been reported using lasers
of different wavelengths and pulse durations since the 1990's
\cite{Paulus_PRL94, Grasbon}. In fact, the plateau in the
high-energy ATI spectra was first observed in Xe. In Fig. 3(a) we
show the calculated ATI spectra vs electron energy in units of $U_p$
at the same laser intensity of $1.0\times10^{14}$ W/cm$^2$ for a
five-cycle, 800 nm pulse. In the case of Xe, the plateau is clearly
seen -- it remains almost constant from  $5U_p$ to $10U_p$, while
for Kr and Ar, each drops about two to three orders of magnitude in
the same energy range. Since the momentum distributions of the
returning electron wave packets are essentially identical (up to a
normalization due to the different tunneling ionization rates) for
the same laser pulse, as shown in Fig.~3(b), the differences in the
electron energy spectra are attributed to differences in the DCS at
large angles among the three targets. In Figs. 3(d-f) their DCS are
shown for the relevant range of $p_r$. Note that they cover the same
range of magnitude, but at large angles, say
160$^\circ$-180$^\circ$, the DCS for Xe behaves ``anomalously'' for
it increases with increasing energies instead of otherwise, as in Ar
and Kr. Since higher returning energies contribute more to the
plateau electron yields (see Fig.~1), this ``anomalous'' energy
dependence   explains the flat plateau in Xe. Likewise, the
``normal'' energy dependence of DCS in Ar and Kr can be used to
explain the steep drop of their electron spectra at high energies,
see Fig.~3(a). We comment that the QRS results shown in Fig.~3(a)
are again in good agreement with those obtained from TDSE.

According to the QRS model, for different lasers with identical
$U_p$, their returning electron wave packets will have the same
range of kinetic energy, thus similar plateau electron momentum and
energy distributions. In Fig.~3(c), we show the electron spectra for
Xe using 800 nm and 400 nm lasers, respectively, but with the latter
having four times the peak intensity. The spectra in the plateau
region indeed agree quite well quantitatively after they are
normalized to each other at around $10U_p$. Both results are
obtained from solving the TDSE.

We also comment that the QRS model is based on the rescattering
theory, thus it would fail at lower intensities in the multiphoton
ionization regime. In Fig.~3(c) we show the comparison of results
from TDSE and from QRS at two lower laser intensities, with Keldysh
parameters of $\gamma=1.59$ and 2.24, respectively. We see clear
evidence of the failure of the QRS model at $\gamma=2.24$. This sets
a rough upper limit for the validity of QRS at $\gamma$ near about
2.0.

The above examples illustrate that the behavior of laser-induced
plateau electrons with energies from $4U_p$ to $10U_p$ are
determined by electron-ion elastic scattering cross sections for
$\bf {free}$ electrons with energies between $1.24U_p$ and $3.17U_p$
(or momentum between 0.79-$1.26A_0$). Flat plateau is expected when
the DCS at large angles (close to 180$^\circ$) increases with
increasing kinetic energy and when they are highly peaked at large
angles, as in the case of Xe [Fig.~3(d)]. Such conditions occur
often in the DCS in low energy electron-ion collisions. Where does
it occur depends on electron energies and target species. Based on
the DCS shown in Fig. 3, the plateau will become more pronounced for
Ar and Kr, but less so for Xe, as the laser intensity is decreased.
This is consistent with earlier experimental results \cite{Grasbon}.

In summary we studied laser-induced high-energy plateau electrons.
Together with our previous results where the species dependence of
HHG was traced to their photo-recombination cross sections
\cite{toru08,AT_JPB,AT_prl}, we now have established a quantitative
rescattering (QRS) theory for laser-induced ATI electron and HHG
spectra in the plateau region. Based on the QRS model which is valid
in the plateau region, there is no need to solve TDSE directly. For
HHG one only needs to calculate photo-recombination cross sections
and for plateau electrons one needs only to calculate elastic
electron-ion scattering cross sections. The returning electron wave
packets, as shown previously \cite{chen07,AT_prl}, can be extracted
from a companion target or from the second-order strong-field
approximation. Conversely, the QRS theory allows one to extract
electron and photon scattering information from laser-induced
electron spectra or HHG, respectively. As proposed elsewhere
\cite{toru_NJP}, these cross sections can be further used to deduce
the structure of the target, thus opening up the opportunity of
using infrared laser pulses for determining the structural change of
a dynamic system with temporal resolution of sub-femtoseconds to a
few femtoseconds.

\section{Acknowledgment}

This work was supported in part by Chemical Sciences, Geosciences
and Biosciences Division, Office of Basic Energy Sciences, Office of
Science, US Department of Energy. TM is also supported by a
Grant-in-Aid for Scientific Research (C) from MEXT, Japan, and by a
JSPS Bilateral joint program between US and Japan.

\end{document}